\documentclass[lettersize,journal]{IEEEtran}

\usepackage[]{soul, color, xcolor}

\usepackage{amsmath,amsfonts}
\usepackage{algorithmic}
\usepackage{algorithm}
\usepackage{array}
\usepackage[caption=false,font=normalsize,labelfont=sf,textfont=sf]{subfig}
\usepackage{textcomp}
\usepackage{stfloats}
\usepackage{url}
\usepackage{verbatim}
\usepackage{graphicx}
\usepackage{cite}
\usepackage{booktabs}

\usepackage{makecell}
\usepackage[colorlinks,linkcolor=black]{hyperref}
\usepackage{cleveref}

\begin{document}

\title{Linear Attention Based Deep Nonlocal Means Filtering for Multiplicative Noise Removal}

\author{
  Siyao Xiao, Libing Huang, Shunsheng Zhang, and Wen-Qin Wang,~\IEEEmembership{Senior Member,~IEEE}
  \thanks{
    Siyao Xiao and Shunsheng Zhang are with the Research Institute of Electronic Science and Technology, University of Electronic Science and Technology of China, Chengdu 611731, China. (email: 202222230113@std.uestc.edu.cn, zhangss@uestc.edu.cn)
  }
  \thanks{
    Libing Huang and Wen-Qin Wang are with the School of Information and Communication Engineering, University of Electronic Science and Technology of China, Chengdu 611731, P.R.China. (email: huanglibing94@foxmail.com, wqwang@uestc.edu.cn)
  }
}

\markboth{Journal of \LaTeX\ Class Files,~Vol.~14, No.~8, August~2021}%
{Shell \MakeLowercase{\textit{et al.}}: A Sample Article Using IEEEtran.cls for IEEE Journals}


\maketitle

\begin{abstract}
  Multiplicative noise widely exists in radar images, medical images and other important fields' images. Compared to normal noises, multiplicative noise has a generally stronger effect on the visual expression of images. Aiming at the denoising problem of multiplicative noise, we linearize the nonlocal means algorithm with deep learning and propose a linear attention mechanism based deep nonlocal means filtering (LDNLM). Starting from the traditional nonlocal means filtering, we employ deep channel convolution neural networks to extract the information of the neighborhood matrix and obtain representation vectors of every pixel. Then we replace the similarity calculation and weighted averaging processes with the inner operations of the attention mechanism. To reduce the computational overhead, through the formula of similarity calculation and weighted averaging, we derive a nonlocal filter with linear complexity. Experiments on both simulated and real multiplicative noise demonstrate that the LDNLM is more competitive compared with the state-of-the-art methods. Additionally, we prove that the LDNLM possesses interpretability close to traditional NLM. The source code and pre-trained model are available at \url{https://github.com/ShowiBin/LDNLM}.

\end{abstract}

\begin{IEEEkeywords}
  Image Despeckling, Deep Learning, Multiplicative noise, Synthetic Aperture Radar, Nonlocal Means Filtering.
\end{IEEEkeywords}

\section{Introduction}

\IEEEPARstart{M}{ultiplicative} noise is also known as speckle, a common noise among active imaging systems, like ultrasound imaging and synthetic aperture radar (SAR) imaging. This kind of noise affects the works based on the imaging results seriously. Unlike additive noise, the visual effect of multiplicative noise is usually more severe. Meanwhile, the coherent process of the active imaging system makes the multiplicative noise inevitable, therefore it is difficult to obtain clean references for training. The removal of multiplicative noise can help the application of various imaging systems in remote sensing, medical treatment and so on.

The typical multiplicative noise denoising method contains 3 categories: spatial filtering, transform domain filtering and nonlocal filtering. Spatial filtering \cite{lopes1990maximum, lee1980digital} mainly works by weighted averaging the pixels in the local range, and the weights can be set for different purposes. Transform domain filtering \cite{xie2002sar, achim2003sar} mainly works by transforming the origin image to another domain to separate the noise signal and the signal of interest. The images are usually transformed into the frequency domain or wavelet domain. The nonlocal filtering \cite{buades2005non, buades2011non, parrilli2011nonlocal} mainly works by weighted averaging the pixels in nonlocal range. The weights can be obtained by the similarity calculation of neighborhood matrices.

Recently, numerous studies have been conducted on deep learning-based denoising methods to solve the image denoising problem. Inspired by the success of convolution neural networks (CNN), \cite{zhang2017beyond, yue2020dual} construct denoising networks with CNN and continuously optimize the network structure. \cite{joo2019dopamine, chierchia2017sar, cheong2021oct} introduce the CNN to multiplicative noise removal, and achieved superior performance. These methods usually utilize deep neural networks to adjust their parameters through training, and then the trained network is adopted as a non-linear function to convert a noisy image into a filtered output. Considering the unique feature of multiplicative noise, \cite{wang2017sar} proposed a neural network to predict the noise of each pixel in the image and then divide the original image and noise at the pixel level to obtain the filtered result. Based on the diffusion model \cite{diffusion}, \cite{DDPM1} and \cite{DDPM2} employ an autoregressive paradigm on multiplicative noise removal.  These methods mostly focus the denoising using the local information but do not pay much attention to the nonlocal information. Based on the attention mechanism \cite{attention} and vision transformer (ViT) \cite{dosovitskiy2020image}, many works \cite{3dAuto, li2023spectral, lai2023hybrid, yang2022low,swinIR, TransSAR} attempt to apply the transformer-like architectures to image restoration. However, the performance of these supervised methods still suffers from their huge network architecture and high complexity.

Due to the lack of clean references for training, the speckled image synthesis method is usually adopted to prepare the training pairs \cite{chierchia2017sar}. Some work \cite{ma2020sar, vitale2021analysis} aimed at constructing references with complementary images. Besides, researchers attempt to overcome the shorts of training data synthesizing from other prospectives, such as proposing loss that takes care of spatial and statistical properties \cite{joo2019dopamine, vitale2020multi}. Inspired by the unsupervised methods of natural image denoising \cite{lehtinen2018noise2noise, han2023ss,wang2023noise2info}, researchers construct unsupervised methods using temporal diversity \cite{dalsasso2021sar2sar}, spatial diversity \cite{molini2021speckle2void} and diversity of real and imaginary parts \cite{dalsasso2021if}. However, the strict assumptions of these unsupervised methods make there room for further improvement.

Additionally, since the deep neural network is only a black box in the denoising model, for the above-mentioned deep learning-based image denoising methods, it is difficult to explain how the results of them are produced. To address the above problem, \cite{lefkimmiatis2017non}, \cite{cozzolino2020nonlocal}, \cite{denis2019patches} and \cite{TransNLM} attempted to incorporates the deep models with NLM, but the interpretability of them is still limited. Unlike them, we employ deep channel CNN to extract the pixels' semantics first. Then, the inner product of the extracted vectors is adopted to calculate the similarities of pixels. Finally, the weighted averaging of extracted vectors and a linear projection are performed to obtain the filtered results. Additionally, the kernel functions are used to make the similarity calculation linearized. The main contributions of this work can be summarized as follows:

\begin{itemize}
  \item We proposed a new denoising method LDNLM to optimize the NLM with deep channel CNN and kernel function based linear attention, resulting in a more interpretable and more efficient deep nonlocal means denoising method.
  \item Through changing the calculation order of inner vectors, we derive a nonlocal means denoising algorithm with linear complexity.
  \item We discuss the role of individual modules in LDNLM with ablation experiments and discuss the interpretability of the LDNLM with visualization.
\end{itemize}

\section{RELATED WORK}
\subsection{Nonlocal Denoising}
The nonlocal methods search pixels similar to the pixel to be processed on a large range and obtain the filtered result by weighted averaging of them.

The traditional local filtering can be formulated as follows:

\begin{equation}
  \hat{x}=\mathbf{A}^{\circ} \mathbf{\Omega},
\end{equation}

\noindent where $\mathbf{A}$ is the convolution kernel, $\mathbf{\Omega}$ refers to the patch centered on the pixel to be denoised. $\mathbf{\Omega}$ is represented in matrix form, which is called the search window here. Essentially, this operation can be understood as the weighted averaging of pixels in $\mathbf{\Omega}$.

\begin{equation}
  \hat{x}=\sum_{a \in \mathbf{A}, x \in \mathbf{\Omega}} a\times x.
\end{equation}

On this basis, nonlocal filtering expands the search window $\mathbf{\Omega}$, computes the weights based on the similarity calculations and then obtains filtered results by weighted averaging.

In many situations, the noise in the images is serious and the noise model is not completely subject to a particular distribution. It’s impossible to restore the original pixel based on only the local pixels, while nonlocal filtering can find pixels relatively cleaner from a large search area to supplement the denoising. As long as the search area is large enough and there is a probability of finding similar pixels, the nonlocal method is capable of performing well.

Based on this idea, NLM \cite{buades2011non} extracts the pixel information with the neighborhood matrices and calculates the similarities using the Euclidean distance. BM3D\cite{dabov2007image} takes the image patches as prediction units and combines the 3D transform and wiener filtering to improve the denoising performance.

Although these nonlocal filtering methods improve denoising performance, they lack a flexible similarity calculation and a faster inference speed. We employ neural networks to replace the key steps inside the nonlocal filtering, and then linearize the similarity calculation with kernel-based transforming.

\subsection{Interpretability}

Interpretable AI deals with developing explainable models. Scholars defined interpretability as how an AI agent provides reasoning for their decision-making. This makes the system more reliable and trustworthy \cite{kaur2022trustworthy}.

Multiplicative noise usually exists in radar images and medical images. The denoising of these images helps the downstream tasks such as radar target tracking, and medical diagnosis. The application scenarios of these tasks are mostly critical situations where a small change can have a significant impact. In these scenarios, interpretability in decision-making is critical.

Generally, traditional denoising methods embody strong interpretability, such as NLM. They have a clear structure and rigorous mathematical derivation. However, deep learning based methods usually treat neural networks as black boxes, where the removal of noise is entirely predicted based on the statistical characteristics of the training data. Therefore, Lefkimmiatis et al. \cite{lefkimmiatis2017non} incorporated the NLM algorithm into CNN, which uses a deep CNN to replace the similarity computation in the NLM algorithm. They use the image information in the search window as the model input, and calculate the model output by exploiting the similarity between neighborhood matrices. The final pixel prediction is calculated with the weight values. Cozzolino et al. \cite{cozzolino2020nonlocal} and Denis et al. \cite{denis2019patches} introduced the above idea to multiplicative noise removal, but these methods still have low interpretability as there are many components lacking clear operational logic.

We enhance the key steps in traditional nonlocal filtering, and optimize the pixel information representation step and similarity calculation step using deep networks. Additionally, we use the kernel function to speed the inference and improve the performance. By doing so, the proposed LDNLM has a better performance and a faster inference speed while the rigorous logical derivation is also highly retained.

\section{METHODOLOGY}

In this section, we propose a linear attention mechanism based on deep nonlocal means filtering (LDNLM). On the basis of traditional NLM, we first use the deep channel CNN to extract the pixels' neighborhood information, mapping the pixels to high dimension space. Then we further linear map them to Query, Key and Value vectors. After that, we employ the inner product and weighted averaging of multi-head attention instead of the similarity calculation and weighted averaging of NLM. Finally, we replace the similarity calculation with kernel mapping, and change the calculating order of inner product and weighted averaging, making the complexity of nonlocal denoising reduce to linear.

\subsection{Deep Nonlocal Means Filtering}

\begin{figure*}[tb]
  \centering
  \includegraphics[width=15cm]{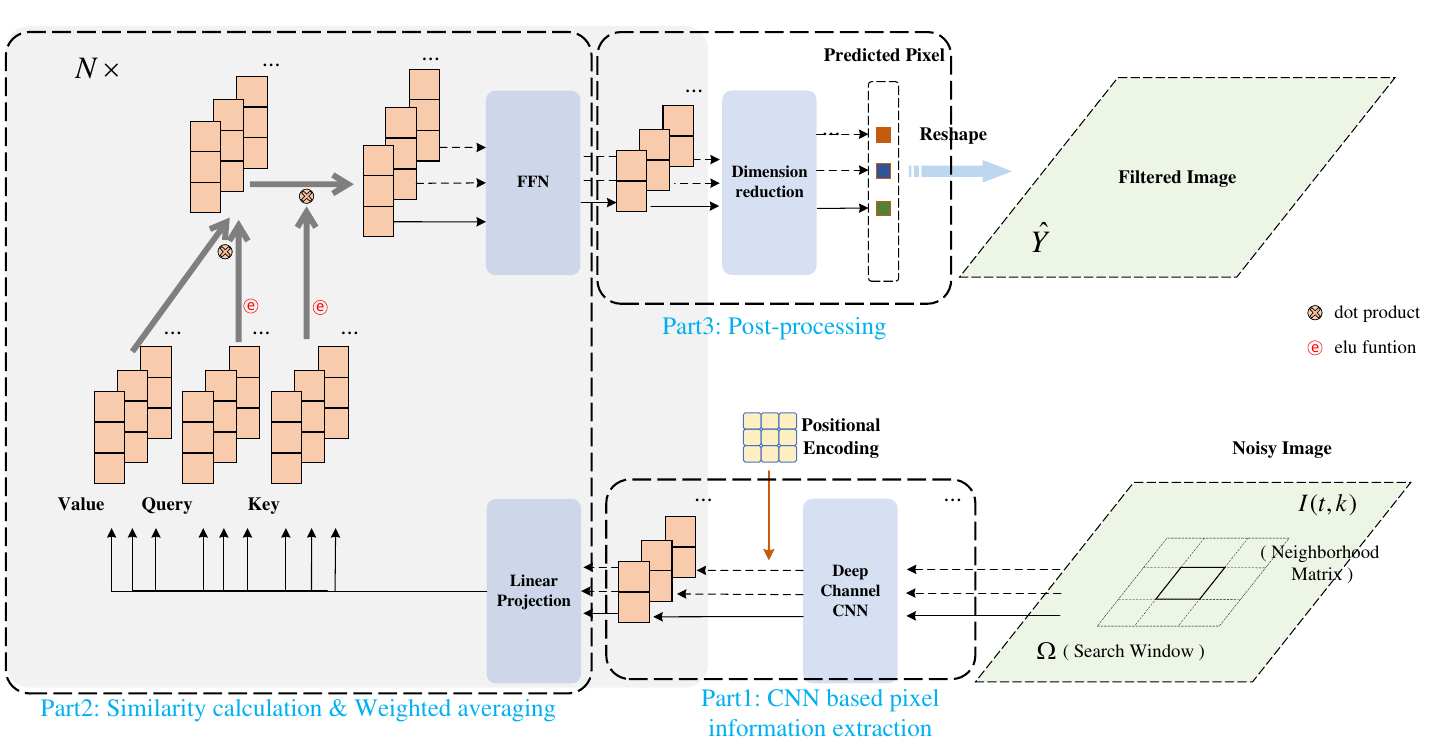}
  \caption{The basic framework of LDNLM. The whole framework contains three parts: CNN based pixel information extraction, Similarity calculation \& Weighted averaging and Post-processing.
  }
  \label{framework}
\end{figure*}

The framework of LDNLM can be draw as \cref{framework}. Assuming in the search window $\mathbf{\Omega}$ in a noisy image, the $\textbf{I}(t,k)$ refers to a neighborhood matrix centered at $t$ with size of $(2k+1)\times (2k+1)$. For extracting the geometrical information, we employ $d$ CNNs $\mathbf{M}^{\theta}$ on $\textbf{I}(t,k)$, the output of $i$-th CNN $\mathbf{M}_{i}^{\theta}$ can be formulated as:

\begin{equation}
  \mathbf{I}_{\mathbf{i}}^{\prime}(t, k)=\max \left(0, \mathbf{M}_{i}^{\theta} \odot \mathbf{I}(t, k)\right),
\end{equation}

\noindent where the $\odot$ is element-wise multiplication.

For following operations, the obtained $\mathbf{I^{\prime}}(t, k)$ must be flattened. Meanwhile, the positional encodings are added to $\mathbf{I^{\prime}}(t, k)$ to provide the prior information about the order of the data. The value of the $j$-th dimension in the position encoding located at position $i$ in $\mathbf{I^{\prime}}(t, k)$ is computed using the following equation.

\begin{equation}
  \left\{\begin{array}{l}
    \mathbf{P}_{(i, 2 j)}=\sin \left(i / 10000^{2 j / d}\right) \\
    \mathbf{P}_{(i, 2 j+1)}=\cos \left(i / 10000^{2 j / d}\right)
  \end{array}\right. .
\end{equation}

Then, $\mathbf{I^{\prime}}(t, k)$ is mapped to high-dimension vectors Query $\textbf{Q}$, Key $\textbf{K}$ and Values $\textbf{V}$. This makes the similarity calculation more accurate for the specific domain.

\begin{equation}
  \label{positional_encoding}
  \left\{\begin{array}{l}
    \textbf{K}=\textbf{W}_K \times \textbf{I}_{v}(\textbf{X}, k)+\textbf{B}_K \\
    \textbf{Q}=\textbf{W}_Q \times \textbf{I}_{v}(\textbf{X}, k)+\textbf{B}_Q \\
    \textbf{V}=\textbf{W}_V \times \textbf{I}_{v}(\textbf{X}, k)+\textbf{B}_V
  \end{array}\right. ,
\end{equation}

\noindent where $\textbf{W}_Q,\textbf{W}_K,\textbf{W}_V$ and $\textbf{B}_Q,\textbf{B}_K,\textbf{B}_V$ are randomly initialized matrices. Here, we use a high-dimension representation to replace the neighborhood matrices in NLM. The $V$ corresponds to the pixels at certain positions in the NLM and is used for the weighted averaging to obtain the target values.

Tradition NLM computes the similarity with Euclidean distance. For effective computing, we use the inner product to calculate the similarities. Borrow the expression from the multi-head attention mechanism, the weights obtained through attention calculating can be represented as:

\begin{equation}
  w= \text{softmax}(\textbf{Q}\textbf{K}^T/\sqrt{d_k}),
\end{equation}

\noindent where $d_k$ is the dimension size of vectors in $\textbf{Q}$, $\textbf{K}$ or $\textbf{V}$.

Therefore, the representation vectors of pixel prediction can be calculated as follows:

\begin{equation}
  \label{Attention}
  \text{Attention}(\textbf{Q}, \textbf{K}, \textbf{V})=\operatorname{softmax}\left(\frac{\textbf{Q} \textbf{K}^{T}}{\sqrt{d_{k}}}\right)\textbf{V},
\end{equation}

After that, the representation vectors are fed to a feedforward neural network (FFN) to extract a more non-linear representation of features.

\begin{equation}
  \label{FFN}
  \text{FFN}(x)=\max \left(0, x \textbf{W}_{1}+\textbf{B}_{1}\right) \textbf{W}_{2}+\textbf{B}_{2},
\end{equation}
\noindent where $\textbf{W}_1, \textbf{W}_2$ and $\textbf{B}_1,\textbf{B}_2$ are randomly initialized matrices.

Additionally, layer normalization and residual learning are both performed after the attention calculation and the FFN calculation, respectively, which makes the whole network easier to train.

Additionally, the dimensions of the vectors obtained through the processing of \cref{positional_encoding} (i.e., position encoding) and \cref{FFN} (i.e., FFN), are the same. This allows the part between to be treated as a cell. Like Transformers, we can stack multiple cells to get an improved performance without modifying the model details.

After the calculations, we obtain an vector representation of the final pixel prediction of each position. Finally, with the vector representation, we can perform dimension reduction to get a pixel value, i.e., the final prediction.

\subsection{Linear Attention Based Deep Nonlocal Means Filtering}

Whether the traditional nonlocal means filtering, or the above deep nonlocal means filtering, both suffer from low inference speed and high memory usage. They need to calculate the similarities of every pixel with each other, the complexity of this whole process is $O(n^2)$. Theoretically, the nonlocal filtering is able to find a pixel similar to the pixel to be processed in the search window, and is possible to get a good denoising result. However, the quadratic complexity limits the search window to be so large, that this unfortunately limits the performance.

The fundamental reason for the quadratic complexity lies in the \cref{Attention}. Where all $N$ pixels in the search window should be used to calculate similarity with all other $N-1$ pixels. For the resulting vector $V_i^{\prime}$ of $i$-th pixel, we can write a generalized attention equation for any similarity function as follows:

\begin{equation}
  V_{i}^{\prime}=\frac{\sum_{j=1}^{N} \text{sim}\left(Q_{i}, K_{j}\right) V_{j}}{\sum_{j=1}^{N} \text{sim}\left(Q_{i},K_{j}\right)} .
  \label{eq7}
\end{equation}

If the $\text{sim}(Q_i, K_j)$ is selected as $\text{exp}(\frac{{Q_i^T} K_j}{\sqrt{d_k}})$, the \cref{eq7} is equivalent to \cref{Attention}.

Note that in order for \cref{eq7} to define an attention function, the only constraint we need to impose to $\text{sim}(\cdot,\cdot)$ is to be non-negative. This includes all kernels:

\begin{equation}
  k(x, y): R^{2 \times d_{k}} \rightarrow R+,
  \label{eq8}
\end{equation}

\noindent where the kernel function $k(x,y)$ also can be formulated as $k(x,y)=\phi(x)\phi (y)$. Given the kernel with feature representation $\phi(\cdot)$ and $\phi(\cdot)$, the \cref{eq7} can be rewritten as follows \cite{katharopoulos2020transformers}:

\begin{equation}
  V_{i}^{\prime}=\frac{\sum_{j=1}^{N} \phi\left(Q_{i}\right)^{T} \phi\left(K_{j}\right) V_{j}}{\sum_{j=1}^{N} \phi\left(Q_{i}\right)^{T} \phi\left(K_{j}\right)} ,
  \label{eq9}
\end{equation}

\begin{equation}
  V_{i}^{\prime}=\frac{\phi\left(Q_{i}\right)^{T} \sum_{j=1}^{N} \phi\left(K_{j}\right) V_{j}^{T}}{\phi\left(Q_{i}\right)^{T} \sum_{j=1}^{N} \phi\left(K_{j}\right)} .
  \label{eq10}
\end{equation}

In summary, the kernel can be a feature map function, employed to calculate similarity. It can be transformed based on the associative property of matrix multiplication.

\begin{equation}
  \left(\phi(Q) \phi(K)^{T}\right) V=\phi(Q)\left(\phi(K)^{T} V\right).
  \label{eq11}
\end{equation}

Because the $ \sum_{j=1}^{N} \phi\left(K_{j}\right) V_{j}^{T}  and  \sum_{j=1}^{N} \phi\left(K_{j}\right) $ can be stored and reused for every query, the transformed operation form has time and memory complexity $O(n)$.

Here, we employ a simple feature map as defined below:

\begin{equation}
  \phi(x)=\text{elu}(x) + 1,
  \label{ea12}
\end{equation}

\noindent where $\text{elu}(\cdot)$ denotes the exponential linear unit \cite{elu} activation function.

\begin{equation}
  \operatorname{elu}(x)=\left\{\begin{array}{c}
    x, x>0 \\
    \alpha\left(e^{x}-1\right), x \leq 0
  \end{array}\right. .
  \label{elu}
\end{equation}

\section{Experiments}

In this section, we first conduct comparative experiments on images containing simulated multiplicative noise. Then, to evaluate the LDNLM’s performance on real images, we conduct comparative experiments on SAR images containing multiplicative gamma noise. After that, the ablation study is performed to analyze the effect of each improvement proposed. Finally, we discuss and validate the interpretability of LDNLM.

All experiments are conducted on an Nvidia RTX 2080Ti GPU with 11GB of video memory, and the deep learning code framework is PyTorch 1.10.1 based on Python 3.9.5

Usually, there is no clear reference corresponding to the real images containing multiplicative noise. Therefore, we simulate the training dataset by synthetic gamma noise on clear optical images. The noisy images can be formulated as:

\begin{equation}
  X = V Y
\end{equation}
\noindent where $X$ is the noisy image, $Y$ is the theoretical clear image, $V$ is the gamma noise and it follows the following distribution:

\begin{equation}
  \label{deqn_ex1a}
  p(V)=\frac{L^{L} V^{L-1} e^{-L V}}{\Gamma(L)}
\end{equation}

\noindent where $\Gamma (\cdot)$ is the gamma function, and $E(V)=D(V)=1/L$.

Overall, we synthesize the training data by matrix multiplication of clear optical images with gamma-distributed random variables at the pixel level. Here, $L$ is set to 1.

We compare the proposed method with traditional methods and state-of-the-art deep learning-based methods in the experiments. Specifically, the methods for comparison include non-learning based methods nonlocal means \cite{buades2011non}, BM3D\cite{dabov2007image}, supervised deep learning method SAR-CNN\cite{wang2017sar}, MONet \cite{vitale2020multi}, SAR-CAM\cite{ko2021sar}, CNN-NLM \cite{cozzolino2020nonlocal} and Trans-SAR\cite{TransSAR}, unnsupervised deep learning method in \cite{SARDDPM}.

We set the radius of the search window of LDNLM to 36, the neighborhood radius to 9, the number of layers to 2 and the number of heads to 8. Meanwhile, we replicate all supervised denoising methods using the open-source codes provided by these methods.

\subsection{On Simulation Images}
\label{simulation}

We generate the testing dataset with the same method as synthesizing training data. In this experiment, UC Merced land-use data \cite{bagofword} is utilized to synthesize the noisy images. The data was extracted from large remote sensing images from various areas across the United States. It contains clear optical images from 21 scenarios, and with size of $256\times 256$. We select 8 images from each category to synthesize training data, 1 image to synthesize test data and 1 image to synthesize validation data.

Additionally, Peak noise-to-signal ratio (PSNR) and structure similarity (SSIM) are chosen to quantitative evaluate the denoising performance. PSNR and SSIM evaluate the smoothing degree and the structure preservation degree of the denoising results, respectively.

\begin{figure*}[tb]
  \centering
  \includegraphics[width=16cm]{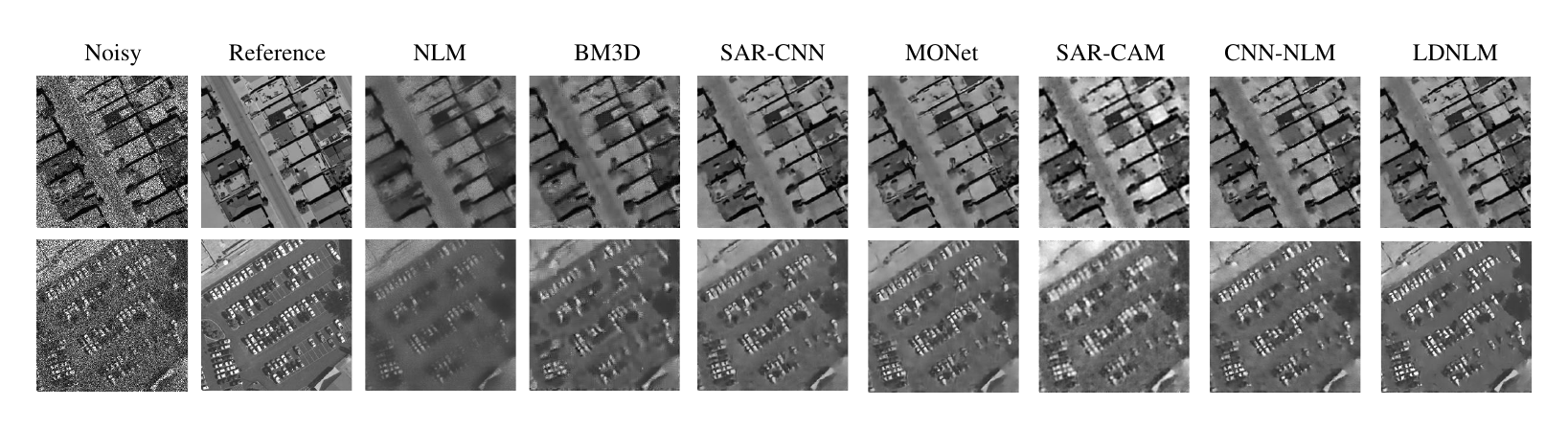}
  \caption{Performance of each model on example simulated images containing multiplicative noise. From left to right: Noisy, Reference, NLM, BM3D, SAR-CNN, MONet, SAR-CAM, CNN-NLM and LDNLM. From top to bottom: dense residential and parking lot.
  }
  \label{sim_exp}
\end{figure*}

The average PSNR and SSIM of each model are as \cref{sim_res}. Additionally, we select NLM, BM3D, SAR-CNN, MONet, SAR-CAM, CNN-NLM and LDNLM, get their denoising results of 2 example simulated images, as \cref{sim_exp}.

\begin{table*}[tb]
  \caption{Average Performance of Each Model on Simulated SAR Images.
  }
  \label{sim_res}
  \centering

  \begin{tabular}{ccc}
    \hline
    Model                                & PSNR            & SSIM           \\
    \hline
    NLM \cite{buades2011non}             & 18.792          & 0.474          \\
    BM3D \cite{dabov2007image}           & 16.820          & 0.447          \\
    SAR-CNN \cite{wang2017sar}           & 24.305          & 0.646          \\
    MONet \cite{vitale2020multi}         & 23.608          & 0.661          \\
    SAR-CAM \cite{ko2021sar}             & 22.843          & 0.595          \\
    Trans-SAR\cite{TransSAR}             & 24.119          & 0.635          \\
    CNN-NLM \cite{cozzolino2020nonlocal} & 23.523          & 0.578          \\
    LDNLM                                & \textbf{25.548} & \textbf{0.695} \\
    \hline
  \end{tabular}
\end{table*}

It can be observed that the LDNLM outperforms other state-of-the-art methods significantly. Traditional methods NLM suffers from over smoothing, BM3D and SAR-CAM are weak in removing the speckles in the single look situation. Deep learning methods SAR-CNN, MONet, CNN-NLM and LDNLM all achieve remarkable denoising performance, while the LDNLM removes the speckles more thoroughly and retains more structure details.

\subsection{On Real SAR Images}

Without loss of generality, we select a type of radar image, namely SAR image, as the noisy real images for denoising experiments. The SAR system uses electromagnetic wave echoes to achieve high-resolution imaging of ground objects. However, due to the principle of coherent imaging, SAR images inevitably suffer from the impact of speckles, i.e., the multiplicative noise.

In this experiment, we employ 2 SAR images from TerraSAR-X. The TerraSAR-X satellite operates in the X-band with a resolution of $3m\times 3m$ in strip mode. We employ single-look images of urban and mountain scenes. The size of the SAR image from the urban scene is $256\times 256$, and the size of the SAR image from the mountain scene is $512\times 512$.

The calculation of PSNR and SSIM requires a clear reference image, however, it is impossible to obtain a clear reference for the image with multiplicative gamma noise. Therefore, we choose the equivalent number of looks (ENL) and the unassisted quantitative evaluation ($\mathcal{M}$) \cite{M_index} to evaluate the performance. ENL is a rough measure of smoothness within a homogeneous region, the higher the better. The ENL of a selected image patch $\bold{P}$ can be calculated as follow:

\begin{equation}
  \mathrm{ENL}=\frac{E\left[\bold{P^2}\right]^{2}}{\operatorname{Var}\left(\bold{P^2}\right)}
\end{equation}
where $E\left[\bold{P}\right]$ and $\operatorname{Var}(\bold{P})$ represent the mean and variance of the image patch $\bold{P}$. The $\mathcal{M}$ measures the visual inspection of the ratio image and the detail preservation capability of a filter. The lower the $\mathcal{M}$, the better the denoising performance.

\begin{equation}
  \mathcal{M}=r_{\widehat{\mathrm{ENL}}, \widehat{\mu}}+\delta h
  \label{M_index}
\end{equation}

As shown in Eq. \eqref{M_index}, $\mathcal{M}$ is the combination of  the $r_{\widehat{\mathrm{ENL}}, \widehat{\mu}}$ and $\delta h$. $r_{\widehat{\mathrm{ENL}}, \widehat{\mu}}$ is the absolute value of the relative residual due to deviations from the ideal ENL, and  $\delta h$ provides an  objective measure for ranking despeckled results within the related ratio images.

\begin{figure*}[tb]
  \centering
  \includegraphics[width=15cm]{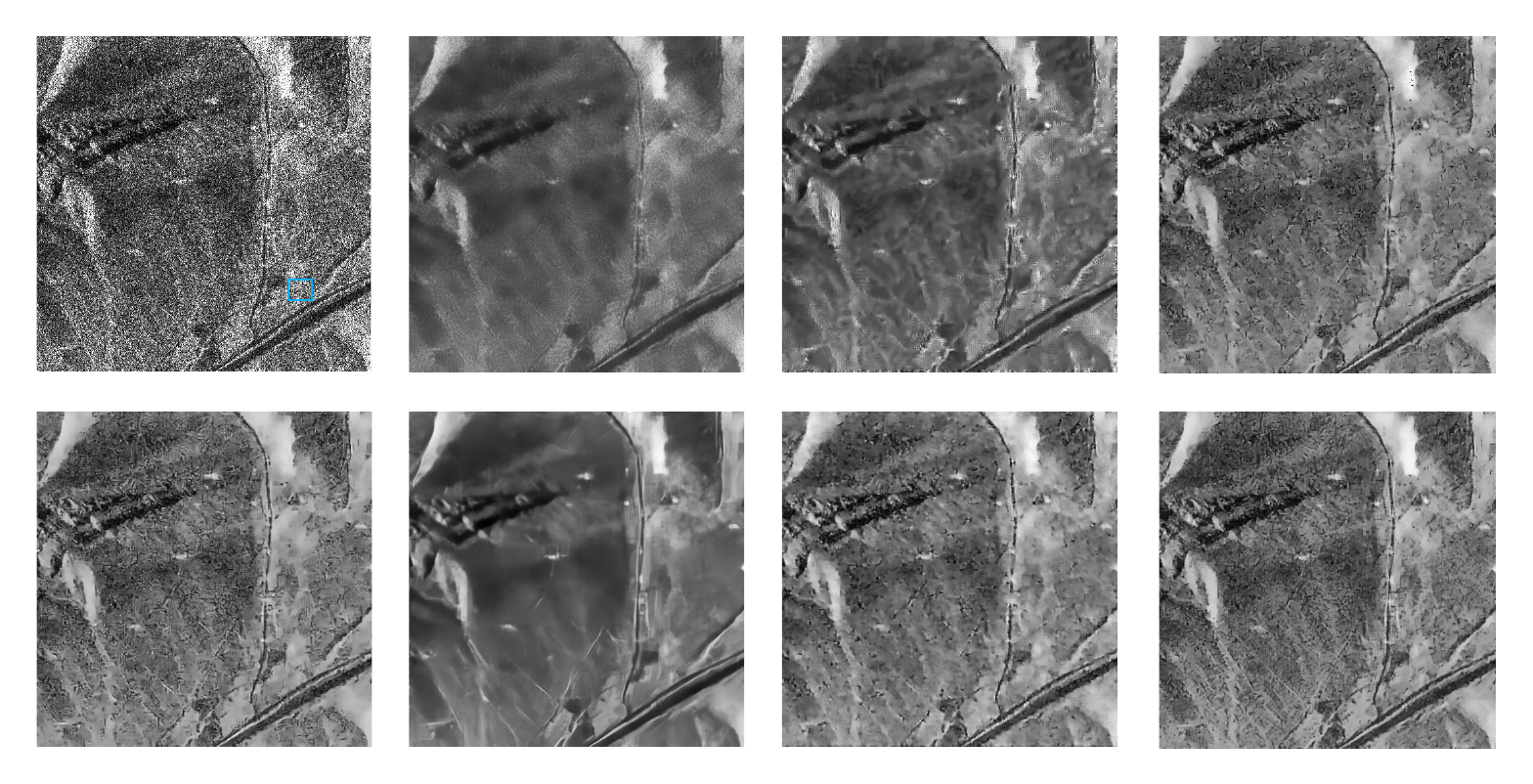}
  \caption{ Performance of each model on real SAR images: TerraSAR-Mountain. From left (top) to right (bottom): Noisy, NLM, BM3D, SAR-CNN, MONet, method in \cite{SARDDPM}, CNN-NLM and LDNLM.
  }
  \label{real_mountain}
\end{figure*}

\begin{figure*}[tb]
  \centering
  \includegraphics[width=15cm]{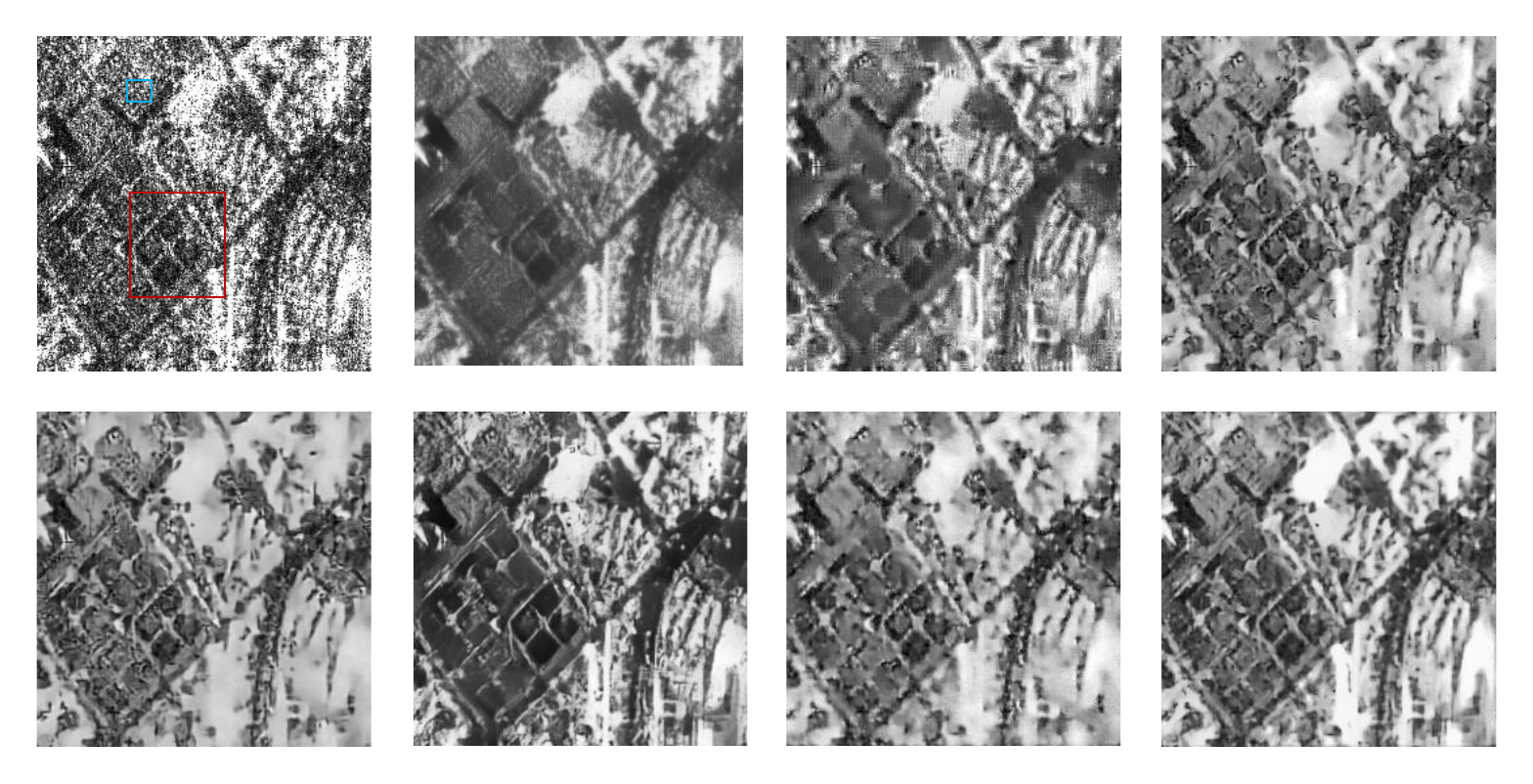}
  \caption{ Performance of each model on real SAR images: TerraSAR-Urban. From left (top) to right (bottom): Noisy, NLM, BM3D, SAR-CNN, MONet, method in \cite{SARDDPM}, CNN-NLM and LDNLM.
  }
  \label{real_urban}
\end{figure*}

\begin{figure*}[tb]
  \centering
  \includegraphics[width=15cm]{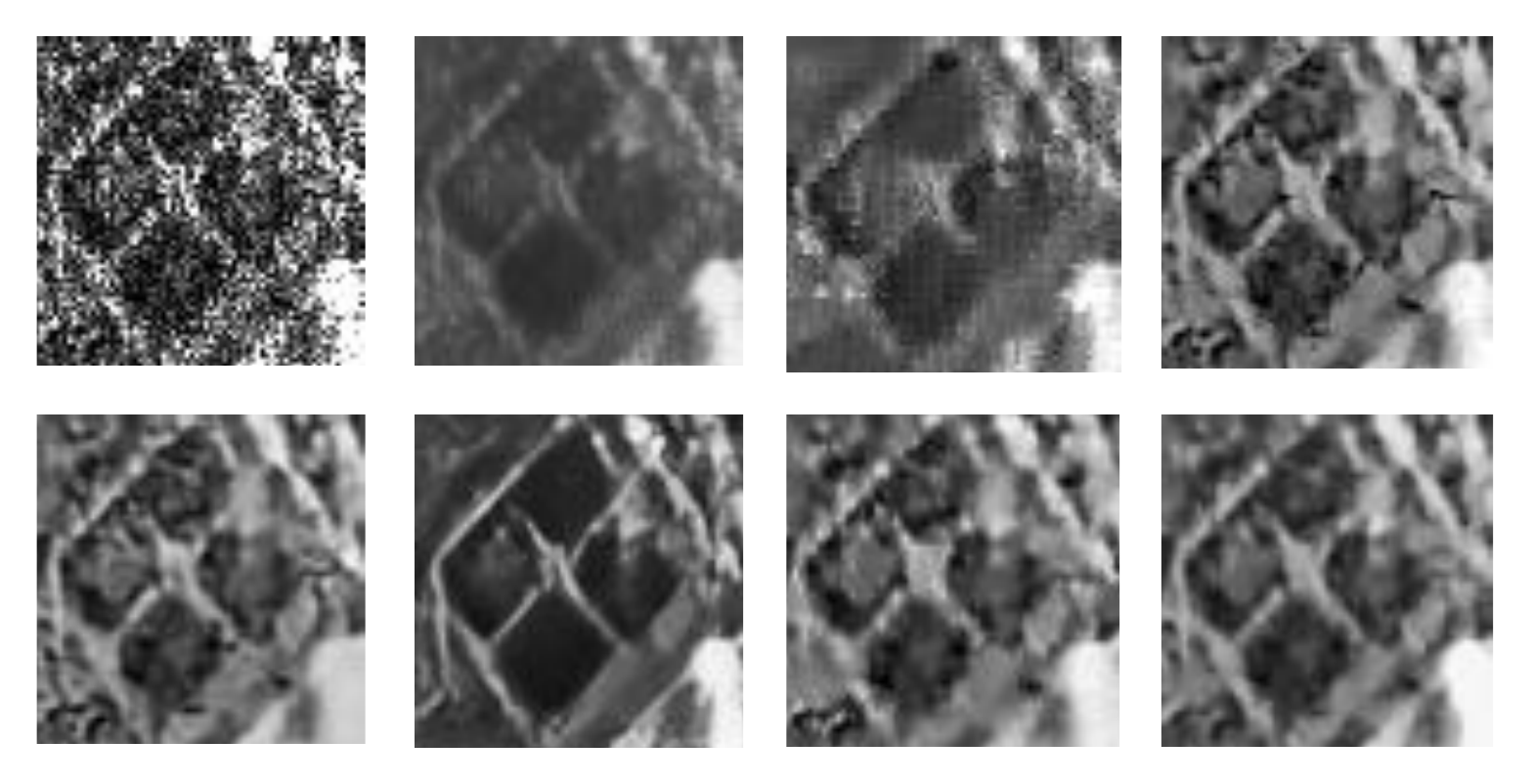}
  \caption{ Performance of each model on real SAR images: The patch indicated by the red boxes in \cref{real_urban}. From left (top) to right (bottom): Noisy, NLM, BM3D, SAR-CNN, MONet, method in \cite{SARDDPM}, CNN-NLM and LDNLM.
  }
  \label{real_mini}
\end{figure*}

\begin{figure*}[tb]
  \centering
  \includegraphics[width=15cm]{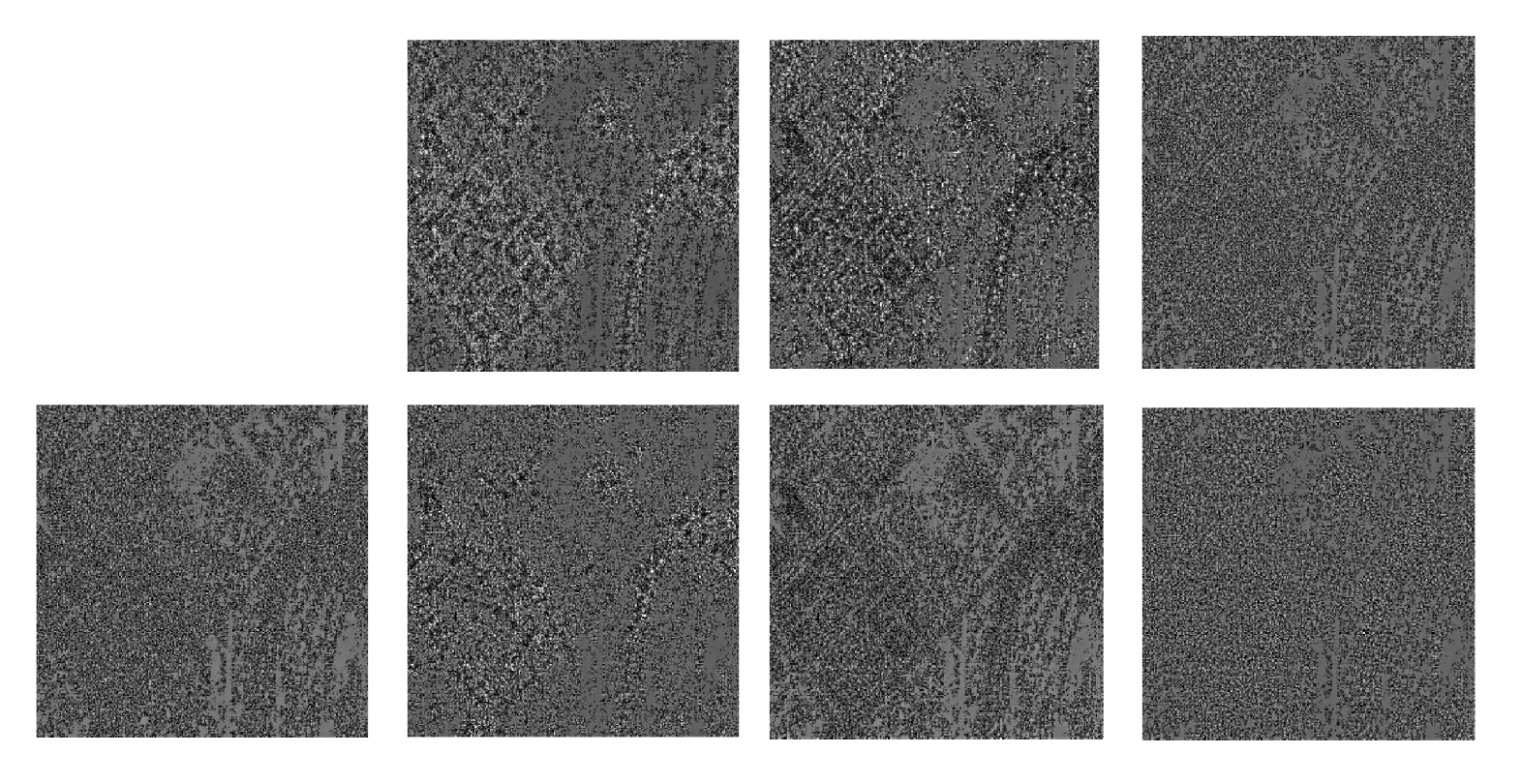}
  \caption{ The ratio image of each model on SAR images: TerraSAR-Urban. From left (top) to right (bottom): NLM, BM3D, SAR-CNN, MONet, method in \cite{SARDDPM}, CNN-NLM and LDNLM. The ratio image follows the statistical properties of the gamma noise ideally.
  }
  \label{ratio_image}
\end{figure*}

To compare the denosing results intuitively, we select NLM, BM3D, SAR-CNN, MONet, method in \cite{SARDDPM}, CNN-NLM and LDNLM, and then get their denoising results of sample SAR images. The selected mountain SAR image and the filtered results of each model are shown in \cref{real_mountain}, and the selected urban SAR image and the filtered results of each model are shown in \cref{real_urban}. The ENL and $\mathcal{M}$ of all filtered results are shown as \cref{real_res}, the blue boxes in the figures indicate the homogeneous regions bounded for the calculation of ENL. To better compare the results of each model, we enlarge the part bounded by the red box in \cref{real_urban} as \cref{real_mini}. Additionally, \cref{ratio_image} shows the ratio images for each model.

\begin{table*}[tb]
  \caption{Performance of Each Model on Real SAR Images. Optimal values are \textbf{bolded} and the second best values are \textit{italicized}.
  }
  \label{real_res}
  \centering

  \begin{tabular}{c c c c c}
    \hline

    Test SAR image                      & \multicolumn{2}{c}{Urban} & \multicolumn{2}{c}{Mountain}                                      \\
    Metrics                             & ENL                       & $\mathcal{M}$                & ENL             & $\mathcal{M}$    \\
    \hline
    NLM\cite{buades2011non}             & 22.147                    & 157.818                      & 27.614          & 46.705           \\
    BM3D\cite{dabov2007image}           & 18.107                    & 136.331                      & 24.838          & 54.233           \\
    SAR-CNN\cite{wang2017sar}           & \textit{23.872}           & {137.495}                    & 39.903          & 61.092           \\
    MONet\cite{vitale2020multi}         & 17.977                    & 176.606                      & \textit{40.058} & \textbf{31.082}  \\
    SAR-CAM\cite{ko2021sar}             & 25.397                    & \textbf{250.260}             & {39.582}        & 66.432           \\
    Trans-SAR\cite{TransSAR}            & 20.795                    & 218.968                      & 42.516          & 96.144           \\
    CNN-NLM\cite{cozzolino2020nonlocal} & 16.864                    & 188.497                      & {26.130}        & 95.6768          \\
    Method in \cite{SARDDPM}            & 12.763                    & 136.309                      & 31.871          & 39.0216          \\
    LDNLM                               & \textbf{25.470}           & \textit{132.255}             & \textbf{42.658} & \textit{38.6902} \\
    \hline
  \end{tabular}
\end{table*}

The ratio image of LDNLM is nearly pure noise, while others' contain many structures of streets. We can observe that the LDNLM gets the relative best texture detail repairing performance in two common SAR scenes. Compared to other methods, LDNLM achieves a competitive noise removal degree. Combing the ratio images and enlarged images of urban scenes, we can find that LDNLM is able to repair structures like roads and buildings whereas SAR-CNN and MONet are doing not good. Unsupervised method in \cite{SARDDPM} achieves a remarkable performance on speckle removal with it's great generating ability, but it does not perform well in texture preservation. It can be concluded that the LDNLM is better at balancing multiplicative noise removal and structure restoration.

\subsection{Ablation Study}

To verify the effectiveness of our strategy, we conduct extensive experiments on the major optimizing components of the proposed LDNLM.

The most important optimizations of LDNLM can be summarized as CNN based pixel information extraction, linear nonlocal denoising and larger search window. Therefore, we conduct comparative experiments on NLM, LDNLM replacing CNN based pixel information extraction with linear projection (LDNLM w/o CNN), LDNLM with standard multi-head attention mechanism (LDNLM w/o linear), LDNLM without CNN based pixel information extraction and linear attention (LDNLM w/o CNN, w/o linear), LDNLM with smaller layer numbers 1 and smaller head numbers 2 (LDNLM-small), LDNLM with larger search window 64, larger neighborhood radius 10 (LDNLM-Large). All models are trained on the synthetic SAR images based on a mini-batch manner, and are ensured to be trained on 560,000 patch samples.

\cref{ablation} shows the results on the simulated noisy image same with \cref{simulation}.

\begin{table*}[tb]
  \caption{Performance of Each Model on Real SAR Images.
    \label{ablation}
  }
  \centering

  \begin{tabular}{cccccccccc}
    \hline
    Model                     & \makecell[c]{Number                                                          \\ of layers} & \makecell[c]{Number \\ of heads} & \makecell[c]{Training time \\ cost(min)} & Batch size & \makecell[c]{Video memory \\ usage(MiB)} & \makecell[c]{Inference \\ time cost(s)} & PSNR & SSIM \\
    \hline
    NLM                       & -                   & - & -   & -  & -              & 1.355 & 18.792 & 0.474 \\
    LDNLM w/o CNN, w/o linear & 1                   & 2 & 197 & 8  & $\approx$10000 & 0.265 & 25.223 & 0.687 \\
    LDNLM w/o CNN             & 1                   & 2 & 73  & 32 & $\approx$6700  & 0.120 & 25.194 & 0.681 \\
    LDNLM w/o linear          & 1                   & 2 & 162 & 8  & $\approx$10300 & 0.256 & 25.279 & 0.683 \\
    LDNLM-small               & 1                   & 2 & 70  & 32 & $\approx$7000  & 0.068 & 25.256 & 0.679 \\
    LDNLM                     & 2                   & 8 & 106 & 32 & $\approx$10600 & 0.088 & 25.548 & 0.695 \\
    LDNLM-Large (64,10)       & 1                   & 2 & 312 & 8  & $\approx$7093  & 0.155 & 25.364 & 0.683 \\
    \hline
  \end{tabular}

\end{table*}

Through the experiment results, we can observe the following three items:
\begin{itemize}
  \item The proposed two optimizations are both able to speed the inference of nonlocal filtering, as well as reduce the memory usage and alleviate the training resources. The substitution of kernel mapping for similarity calculation reduces memory usage and speeds the inference greatly. However, it makes the model performance worse. The deep channel CNN based neighborhood information extracting decreases the training resources to some extent.
  \item These two optimizations are both possible to sacrifice a portion of the model's performance.
  \item Nevertheless, in the same environment with the same memory and closer time cost, by increasing the search window, enlarging the neighborhood matrix, increasing the number of layers and increasing the number of heads, LDNLM is capable of obtaining a superior performance than the original scheme.
\end{itemize}

\subsection{The Interpretability of LDNLM}

Since the LDNLM is an improvement of NLM, the LDNLM remains the same interpretability. The LDNLM can be seen as an improved NLM algorithm, replacing the neighborhood matrix with high-dimension vectors extracted based on CNNs, replacing the similarity calculation with kernel mapping based linear attention, replacing the pixel-wise weighted averaging with vector-wise weighted averaging and adding dimension reduction to predict the final pixel.

While following the NLM idea, the attention mechanism bring the LDNLM a more accurate calculation of similarity. To validate that the linear attention is able to replace the traditional similarity calculation successfully, and can compute the high dimension representation of the filtered results. We obtain the results of attention calculation and interpret them with the visualization.

\begin{figure*}[tb]
  \centering
  \includegraphics[width=8.5cm]{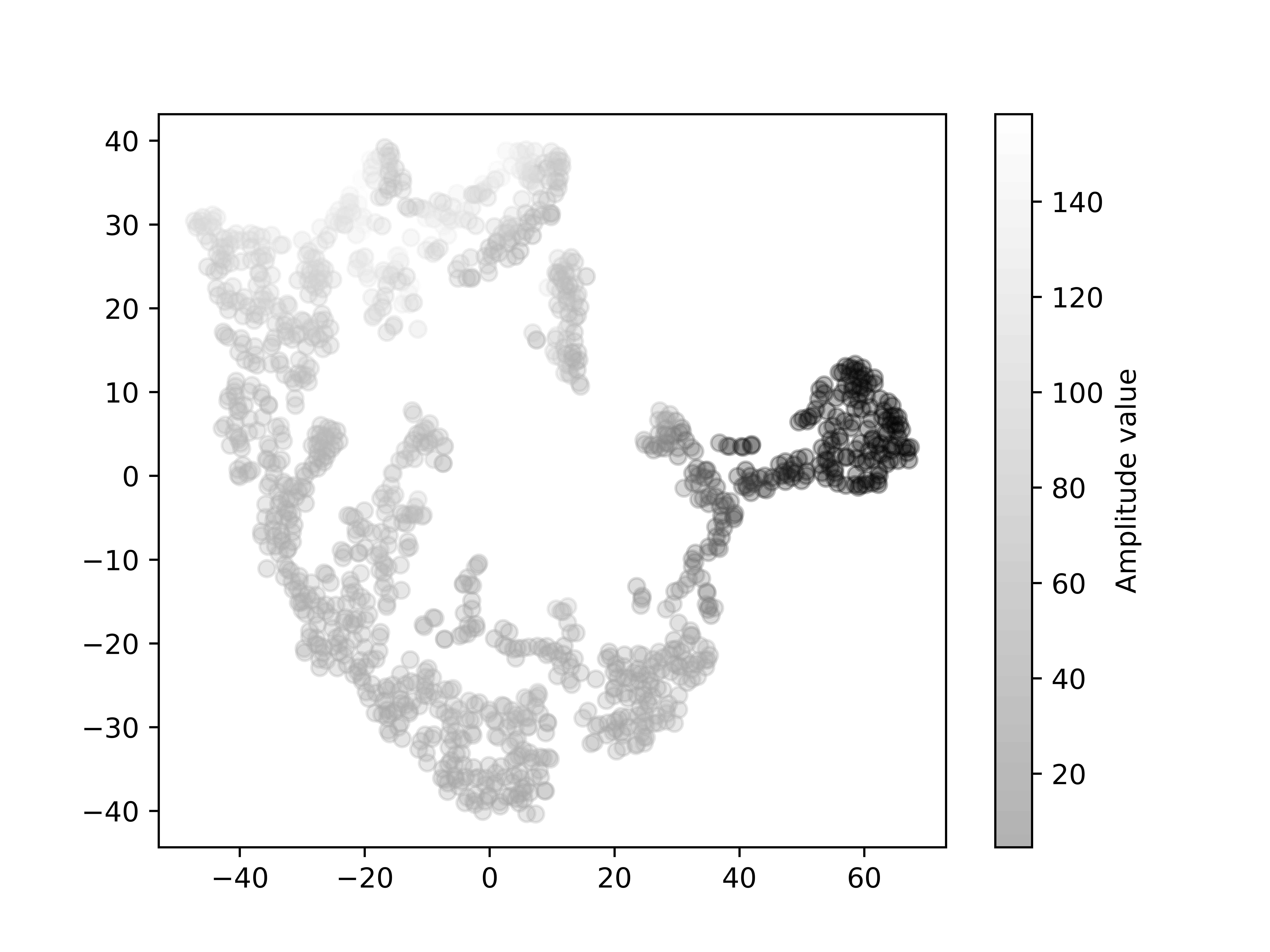}
  \caption{ Dimension reduction and visualization of the representation vectors corresponding to the pixels of the selected area. The dots bounded by the same ellipse can be regarded as one cluster, those reflect the data distribution of the representation vectors.
  }
  \label{visualization}
\end{figure*}

\begin{figure*}[!h]
  \centering
  \includegraphics[width=15cm]{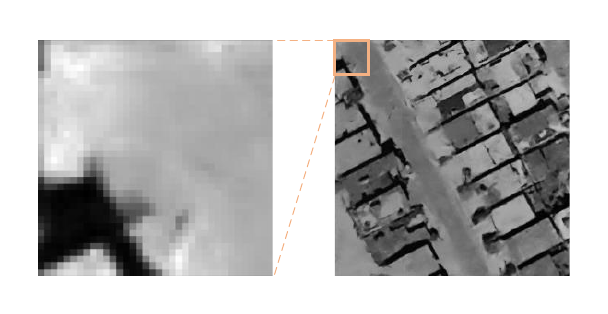}
  \caption{Denoising result of the selected area.
  }
  \label{patch}
\end{figure*}

After the first attention calculation of LDNLM, we get the provisional calculation result. The provisional result theoretically contains $73 \times 73$ vectors, corresponding to all pixels in the search window. For the purpose of making the results easier to read, we firstly extract $36 \times 36$ vectors from all $73 \times 73$ vectors, corresponding to the patch indicated as the yellow box in the filtered image \cref{patch}. The vectors are dimension reduced to 2-dimension based on tSNE \cite{van2008visualizing} as \cref{visualization}. The color represents the final pixels' color in the filtered result, i.e., the amplitude value of synthetic image containing multiplicative noise.

Through the visualization, it can be observed that: The high dimension vectors computed by linear attention are obviously clustered into 2 clusters. The right cluster contains a small number of dots with black color, and the left cluster contains a large number of dots with gray color. Meanwhile, the filtered patch also contains narrow black areas and wide gray areas. The visualization in \cref{visualization} and the filtered result in \cref{patch} correspond to each other.

It's believed that the kernel mapping based linear attention can effectively replace the similarity calculation in nonlocal means filtering.

\section{Conclusion}

In this paper, we optimize the NLM denoising algorithm based on the deep CNN and linear attention, and propose a linear attention based deep nonlocal means filtering for multiplicative noise removal (LDNLM). Inheriting the framework of NLM, we employ deep channel CNN to extract pixel information, and we replace the similarity calculation and weighted averaging part with kernel-based linear attention. These optimizations make the nonlocal algorithm have linear complexity, and improve its performance greatly. Meanwhile, the interpretability of LDNLM is also largely inherited from traditional NLM. In the future, we will explore the LDNLM-based self-supervised strategy for multiplicative noise removal.




%

\bibliographystyle{IEEEtran}
\bibliography{refer}











\newpage

\vfill

\end{document}